\documentclass[11pt]{article}
\usepackage{amsmath}
\usepackage{amssymb}
\usepackage[tmargin = 3cm, lmargin = 3cm, rmargin = 2cm, bmargin = 2cm]{geometry}
\usepackage{graphicx}
    \graphicspath{{./Figures/}}
\usepackage[colorlinks = true]{hyperref}
\usepackage[square, numbers, sort&compress]{natbib}
\usepackage{setspace}
\usepackage{siunitx}
\usepackage{stix2}
\usepackage{subfigure}
\usepackage{xcolor}

\newcommand{\diff}[2]{\frac{\text{d}#1}{\text{d}#2}} % 1st order ordinary derivative \diff{}{}
\newcommand{\hypergeom}[1]{{}_2F_1\left(#1\right)} % hypergeometric function 2F1 \hypergeom{}
\newcommand{\sqabs}[1]{{\left|#1\right|}^2} % square absolute \sqabs{}
\title{{Effective particles in a multishell nanostructure with hardcore}}

\author{
{\normalsize H. R. Christiansen$^{\dagger}$\, {and}\, R. M. Lima$^{*} $ }
\\
\footnotesize{$^{\dagger}$IFCE -\, Instituto Federal de Educação, Ciência e Tecnologia do Ceará, CEP 61940-750, CE, Brazil} 
 \\
\footnotesize{$^{*}$CBPF -\, Centro Brasileiro de Pesquisas Físicas, CEP 22290-180, Rio de Janeiro, RJ, Brazil} 
\\
\footnotesize{corresponding author: hugo.christiansen@ifce.edu.br}
}
\date{}

\begin{document}

\maketitle
\abstract{In-medium effective carriers with position-dependent mass in a multishell heterostructure are analytically studied. We obtain the exact spectrum of three-dimensional bound eigenstates and the scattering wave-functions for several von Roos ordering classes. Ascribing a continuously varying mass to the carriers in a multilayer type spherical system
we use our solutions to compute optical properties such as the absorption coefficients and refraction indices of a nanometric heterostructure.   
We analyze in detail the case of a \(\text{GaAs}/\text{A}\ell\text{GaAs}\) alloy and show how these results depend on the ordering class of the kinetic hamiltonian. 
}
% sec: introduction
\\

\noindent
{\small \textit{Keywords}:  position-dependent mass quantum mechanics; analytical spectrum and eigenstates; quantum heterostructures; quantum dots; nanostructures.}
\\

\noindent
{\small {Published}\footnote{  
https://doi.org/10.1016/j.physb.2024.416564\\ 
https://www.sciencedirect.com/science/article/pii/S0921452624009050} in \textbf{\textit{Physica B: Condensed Matter 695 (2024) 416564}}}

\section{Introduction}
    \label{sec:i}

The modelling of electrons with a mass depending on space coordinates flowing through a quantum system was first applied in solid state physics \cite{wannier:1937, slater:1949, luttinger:kohn:1955, bendaniel:duke:1966, gora:williams:1969}. Instead of the regular constant-mass Schrodinger wavefunction, a so-called effective envelope function was adop-ted to deal with charged carriers in semiconductors \cite{vonroos:1983, bastard:1992}. 
This effective mass approximation can nowadays be understood as a renormalization of the electron mass in response to  the fermion-phonon coupling \cite{whalley:etal:2019}.
Although mathematically involved, the method is satisfactory to accomplish band structure calculations of material features  of semiconductor devices.
For example, transport properties as polaron radii and carrier mobility, defect properties as donor acceptor energy levels, and optical properties as exciton binding energies can be computed from the ground.  

The coupling of electrons to collective excitations is nontrivial so effective approaches are necessary. 
In \cite{zhao:liang:ban:2003} it is shown that there is a clear contribution of the interaction between the electron and the longitudinal optical phonons and the interphase phonons on the polaron energy levels in parabolic quantum wells on the alloy   {\(\text{GaAs}/\text{Al}_\chi\text{Ga}_{1-\chi}\text{As}\)}. A position-dependent effective mass is considered to vary along the axis in the quantum well and is constant in the barrier material.  A  recent model for this kind of system 
is developed in \cite{christiansen:lima:2024}.
Particle-phonon coupling effects have been also discussed in nuclear physics. In \cite{saperstein:etal:2016} the renormalization of the nucleon mass is computed in a many-body approach using a Skyrme Hartree-Fock method.  
Here we will show that both types of physical systems, atomic alloys and nuclear systems, can be treated with the present procedure.

Particles with position-dependent masses (PDM) have been used to address several problems \cite{cunha:christiansen:2013, christiansen:cunha:2013, christiansen:cunha:2014, lima:christiansen:2023a, ho:roy:2019, schmidt:dejesus:2018}. Among these,  
the study of quantum dots and heterostructures in general have been crucial \cite{lima:christiansen:2023b, christiansen:lima:2024,sari:etal:2019, elnabulsi:2020,elnabulsi:2020b, kasapoglu:etal:2021, kasapoglu:duque:2021,valencia-torres:etal:2020, ganguly:etal:2006, galbraith:duggan:1988}.  Quantum dots present advantages like high absorption coefficient,  large intrinsic dipole moment and high photoluminescence quantum yield \cite{selopal:etal:2020}.
The optoelectronic properties of these nanostructures can be fine-tuned by changing their size and shape \cite{talapin:etal:2010, shirasaki:etal:2013, khordad:etal:2011, hassanabadi:rajabi:2009,kasapoglu:etal:2010, atayan:etal:2008} and have a  number of applications in sensors, lasers, biolabels and light emitting diodes  \cite{zeng:etal:2013, vasudevan:etal:2015}.   

Quantum dots have appropriate optical features to control the output of a component device. For instance, one can manage the electronic energy of such structures and adjust the absorption threshold frequency by regulating its growth.
Single-electron transistors can be used to study electron tunneling through a system of  tunnel junctions in series. 
In \cite{kuo:chang:2003}  the electron tunneling current  is studied in a quantum dot irradiated by infrared light within an effective position-dependent mass model. 
Absorption coefficients and relative variations of the refractive index in the inter sub-band transitions between the low-lying  energy levels in  quantum dots have been investigated in regular quantum mechanics for different potentials, see e.g. \cite{wen-fang:2006, mora-ramos:barseghyan:duque:2010, hayrapetyan:kazaryan:tevosyan:2013, hayrapetyan:kazaryan:tevosyan:2012}.  The investigation of the shape effect of the quantum dots under external fields on the electronic spectrum and optical responses plays an important role in semiconductor physics. It helps to simulate real situations and improve the performance of optoelectronic equipments based on low-dimensional heterostructures. Here, we will assume a spherical symmetry. See  \cite{christiansen:lima:2024} for the study of a cylindrical semiconductor nanocrystal with a PDM approach adopting a hyperbolic external potential.

A common deficiency of quantum dots is a high density of surface defects or traps which deteriorate
the overall performance. Fortunately, this flaw can be passivated
through the growth of an outer shell of different material composition. These heterostructures are
called core-shell quantum dots and can enable the broadening
of the absorption spectrum, accelerate the carrier transfer, and reduce exciton
recombination loss \cite{zeng:etal:2013, vasudevan:etal:2015, selopal:etal:2020}.

In the present paper we will compute some optical properties in a PDM approach of a core multishell quantum dot through an analytical calculation of the full spectrum of energy eigenstates.
We will consider a three-dimensional position-dependent mass electronic Hamiltonian with a stepwise external potential. 
In Section \ref{sec:gse} we show the possible PDM kinetic operators and the associated modified Schrodinger equation. In Section \ref{sec:pmd}  the model for the heterostructure is presented. Setting the external potential and the effective position-dependent mass shape we show at full the various differential equations arising for the von Roos ordering classes. The analysis of the boundary conditions specify the exact bound state solutions in all the layers for every kinetic ordering. In Section \ref{sec:rd} we present the applications of this model and include a discussion of scattered waves by the nanostructure. Section \ref{sec:c}  is for the conclusions.

%%%%%%%%%%%%%%%%%%%%%%%%%%%%%%%%%%%%%
% sec: generalized schrödinger equation
\section{Generalized Schrödinger equation} \label{sec:gse}

Quantum mechanics imposes a nontrivial commutation relation between position \(\boldsymbol{{\mathscr{r}}}\) and momentum \(\hat{\boldsymbol{p}} \equiv -i\hbar\boldsymbol{\nabla}\) operators. Thus, the assumption of a PDM turns mass into an operator which does not commute with momenta. This imposes considerations about the order of the (inverse of) mass in the kinetic energy definition. In ref.~\cite{vonroos:1983} von Roos proposed a general form for the kinetic Hamiltonian which parameterizes all the possible Hermitian orderings:
\[\hat{T} = \frac{1}{2{m_0}^2} \hat{\boldsymbol{p}}^2 \longrightarrow \hat{T}_{a,b}(\boldsymbol{{\mathscr{r}}}) = \frac{1}{4} \left( M^a\hat{\boldsymbol{p}}M^{-1 - a - b}\hat{\boldsymbol{p}}M^b + M^b\hat{\boldsymbol{p}}M^{-1 - a - b}\hat{\boldsymbol{p}}M^a\right) \;  .\]
Here, \(M \equiv M(\boldsymbol{{\mathscr{r}}})\) is the mass operator, and \(a,b \in \mathbb{R}\) are the ordering parameters. Popular special cases are BDD (BenDaniel \& Duke, $a = b = 0$), GW (Gora \& Williams, $a = -1$, $b = 0$), ZK (Zhu \& Kroemer, $a = b = -1/2$), LK (Li \& Kuhn, com $a = 0$, $b = -1/2$) and MM (Mustafa \& Mazharimousavi, $a = b = -1/4$) \cite{bendaniel:duke:1966, gora:williams:1969, zhu:kroemer:1983, li:kuhn:1993, mustafa:mazharimousavi:2007}.

Following \cite{lima:christiansen:2023a}, the general differential equation 
%\(\hat{H}_{a,b}(\boldsymbol{\textcolor{blue}{\mathscr{r}}})\Psi(\boldsymbol{\textcolor{blue}{\mathscr{r}}}) = E\Psi(\boldsymbol{{\textcolor{blue}{\mathscr{r}}}})\) 
for {a PDM particle} in an external potential \(V(\boldsymbol{{\mathscr{r}}})\) can be written as
\begin{multline}
    \label{eq:gse} % eq: generalized Schrödinger equation
    -\frac{1}{m}\boldsymbol{\nabla}^2\psi(\boldsymbol{r}) + \frac{1}{m}\frac{\boldsymbol{\nabla}m}{m} \cdot \boldsymbol{\nabla}\psi(\boldsymbol{r}) \\
    + \frac{1}{m}\left[- \frac{a + b}{2}\boldsymbol{\nabla} \cdot \left(\frac{\boldsymbol{\nabla}m}{m}\right) + \left(ab + \frac{a + b}{2}\right){\left(\frac{\boldsymbol{\nabla}m}{m}\right)}^2+ m\Tilde{V}(\boldsymbol{r})\right]\psi(\boldsymbol{r}) = \Tilde{E}\psi(\boldsymbol{r})
\end{multline}
% COMENT: TROQUEI O COMANDO \eqnarray POR \multline; CORRIGI O SINAL DOS TERMOS ADICIONAIS AO POTENCIAL; TROQUEI OS () POR []
in dimensionless {quantities}
%\(a,b \in \mathbb{R}\) are the von Roos ordering parameters \cite{vonroos:1983} related to the order between the momentum operator \(\hat{\boldsymbol{p}} \equiv -i\hbar\boldsymbol{\nabla}\), and the (inverse of the) PDM \(m \equiv m(\boldsymbol{r})\) in the kinetic hamiltonian. In this equation we have used dimensionless mass, wave function, energy and potentials defined by
\(\boldsymbol{r}=\boldsymbol{{\mathscr{r}}}/\epsilon\), \(m(\boldsymbol{r}) = M(\epsilon\boldsymbol{r})/m_0\), \(\psi(\boldsymbol{r}) \equiv \epsilon^{3/2}\Psi(\epsilon\boldsymbol{r})\), \(\Tilde{V}(\boldsymbol{r}) \equiv V(\epsilon\boldsymbol{r})/\mathcal{E}\) and \(\Tilde{E} \equiv E/\mathcal{E}\) (the  parameters $m_0$, \(\epsilon\) and \(\mathcal{E} \equiv \hbar^2/2\epsilon^2m_0\) have mass, length and energy units, respectively). Along with the uncommon second term of eq.~\eqref{eq:gse}, it is worth noting the nontrivial function appearing aside the external potential within brackets. We will call it the kinetic potential,
\[\Tilde{U}_{a,b}(\boldsymbol{r}) \equiv -\frac{1}{m}\left[\frac{a + b}{2}\boldsymbol{\nabla}\cdot\left(\frac{\boldsymbol{\nabla}m}{m}\right) - \left(ab + \frac{a + b}{2}\right){\left(\frac{\boldsymbol{\nabla}m}{m}\right)}^2\right] \;  ,\]
since it results strictly from the relative momentum and mass ordering and is independent of external forces.

%%%%%%%%%%%%%%%%%%%%%%%%%%%%%%
%%%%%%%%%%%%%%%%%%%%%%%%%%%%%%
% sec: potential and mass distribution
\section{The heterostructure model}
    \label{sec:pmd}

Here we analyse a three-dimensional position-dependent mass particle in a spherical finite potential-well with a hard core. This model could  represent several physical scenarios. For example, it could stand for an effective carrier bound in a heterostructure having an insulator in the central region and a finite second wall to jump to a conducting external crystal structure. It could also deal with an effective electron, forbidden to penetrate the atomic nucleus but allowed to be excited out of the atom. It is valid as well for a nucleon onto a saturated nucleus; see {\cite{fuda:1969, downs:ram:1978} and \cite[pp.~103, 104]{krane:1988}} for a discussion of this situation for ordinary constant-mass particles. Hard cores are also considered in the study of classic  \cite{cervantes:benavides:del-rio:2007} and quantum liquids \cite{burkhardt:1968, baker:etal:1982a, baker:etal:1982b}. In \cite{dong:lozada-cassou:2005} it is discussed a simple constant stepwise PDM in the presence of a hard core potential.

\subsection{The effective external potential}

In spherical coordinates, the solutions of eq.~\eqref{eq:gse} can be written as \(\psi(\boldsymbol{r}) = R(r)\Upsilon(\theta,\phi)\), where \(\Upsilon \equiv \Upsilon_\ell^{\mathscr{m}_\ell}\) are spherical harmonics \cite{shankar:2013} with \(\ell,\mathscr{m}_\ell \in \mathbb{Z}\) (\(\ell \geq 0\) and \(\mathscr{m}_\ell = -\ell,...,\ell\)) being respectively the angular and magnetic quantum numbers. The radial wave function $R(r)$ obeys the following modified radial equation
\begin{multline}
    \label{eq:rgse} % eq: radial generalized Schrödinger equation
- \frac{1}{m}{\left(r\,R(r)\right)}'' + \frac{m'}{m^2}{\left(r\,R(r)\right)}' + \\
+ \left\{ - \frac{1}{m} \left[ \frac{a + b}{2}{ \left( \frac{m'}{m} \right) }' - \left( ab + \frac{a + b}{2} \right){ \left( \frac{m'}{m} \right) }^2 + \frac{a + b}{r}\frac{m'}{m} \right] + \Tilde{V}_\ell(r) \right\} rR(r) = \Tilde{E}rR(r) \;,
\end{multline}
% COMENTS: TROQUEI O AMBIENTE EQNARRAY POR MULTLINE; TROQUEI OS () POR {}
%onde
%%\[\Tilde{U}_{a,b}(r) = -\frac{1}{m}\left[\frac{a + b}{2}{\left(\frac{m'}{m}\right)}' - \left(ab + \frac{a + b}{2}\right){\left(\frac{m'}{m}\right)}^2 + \frac{a + b}{r}\frac{m'}{m}\right]\]
%é o potencial cinético e
where
\[\Tilde{V}_\ell(r) \equiv \Tilde{V}(r) - \frac{m'}{m^2r} + \frac{\ell(\ell + 1)}{mr^2}\]
is the effective external potential. For a constant mass it clearly reduces to the addition of the usual centrifugal barrier, as expected. The model is depicted in Fig.~\ref{fig:1}.

We define the radius of the hard core by \(\epsilon\delta_1\) and a finite potential well of depth {\(\mathcal{E}\Tilde{V}_B\)} in a  {spherical} shell of  {thickness} \(\epsilon(\delta_2-\delta_1)\)  {(\(0 < \delta_1 < \delta_2\) are dimensionless)}. Thus we have three regions defined by
\[V( {\mathscr{r}}) =
\begin{cases}
    \infty                  &   {\mathscr{r}} < \epsilon\delta_1    \\
    -\mathcal{E}\Tilde{V}_B &   \epsilon\delta_1 \leq {\mathscr{r}} \leq \epsilon\delta_2   \\
    0                       &   {\mathscr{r}} > \epsilon\delta_2
\end{cases} \; .\]

% sec: dinamics in internal region
\subsection{ {The mass function and the internal shell}}
    \label{sec:dir}

As already stated, the region \(r < \delta_1\) (the hard core) is not accessible to the particle and thus the wave function has to vanish inside: \({{R(r)}|}_{r < \delta_1} = 0\). In the region \(\delta_1 \leq r \leq \delta_2\) (the internal shell) the particle mass varies with position and we will solve eq.~\eqref{eq:rgse} through point canonical transformations. Its solution depends on the ordering parameter and on the angular quantum number: \({{R(r)}|}_{\delta_1 \leq r \leq \delta_2} \equiv R_{a,b;\ell}^\text{in}(r)\). In the region \(r > \delta_2\) (the external {layer}) the mass is uniform and eq.~\eqref{eq:rgse} reduces to a regular Schrödinger equation and so the solutions depend uniquely on the angular quantum number: \({{R(r)}|}_{r > \delta_2} \equiv R_\ell^\text{ext}(r)\).

\begin{figure}
    \subfigure[\label{fig:1a}]{\includegraphics[width = .49\linewidth]{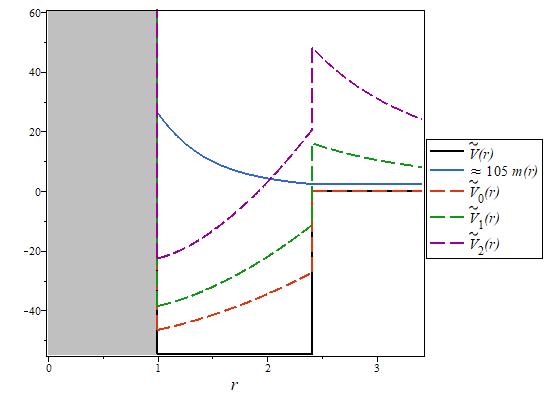}} \ \
    \subfigure[\label{fig:1b}]{\includegraphics[width = .49\linewidth]{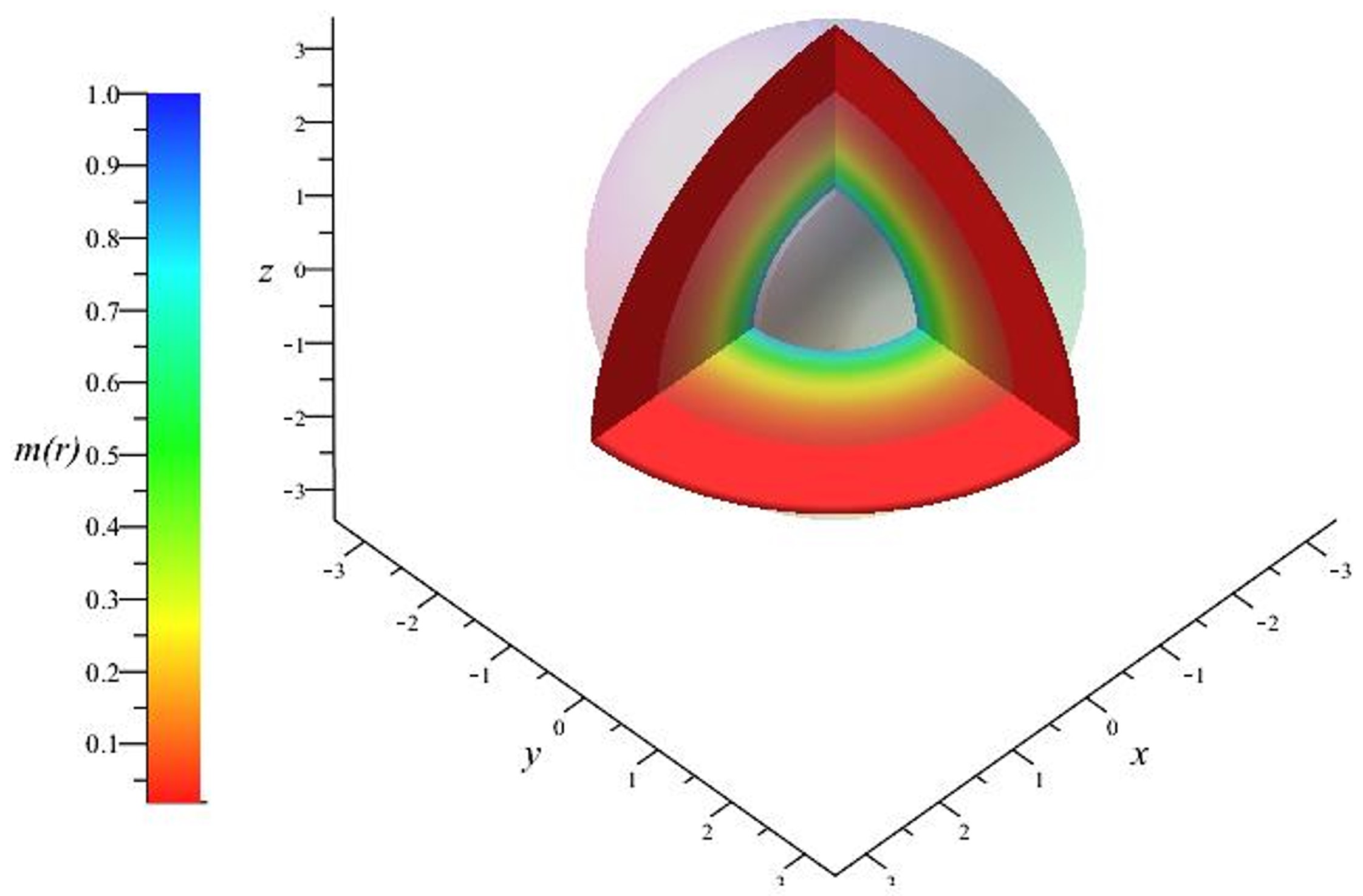}} \\
    \subfigure[\label{fig:1c}]{\includegraphics[width = .32\linewidth]{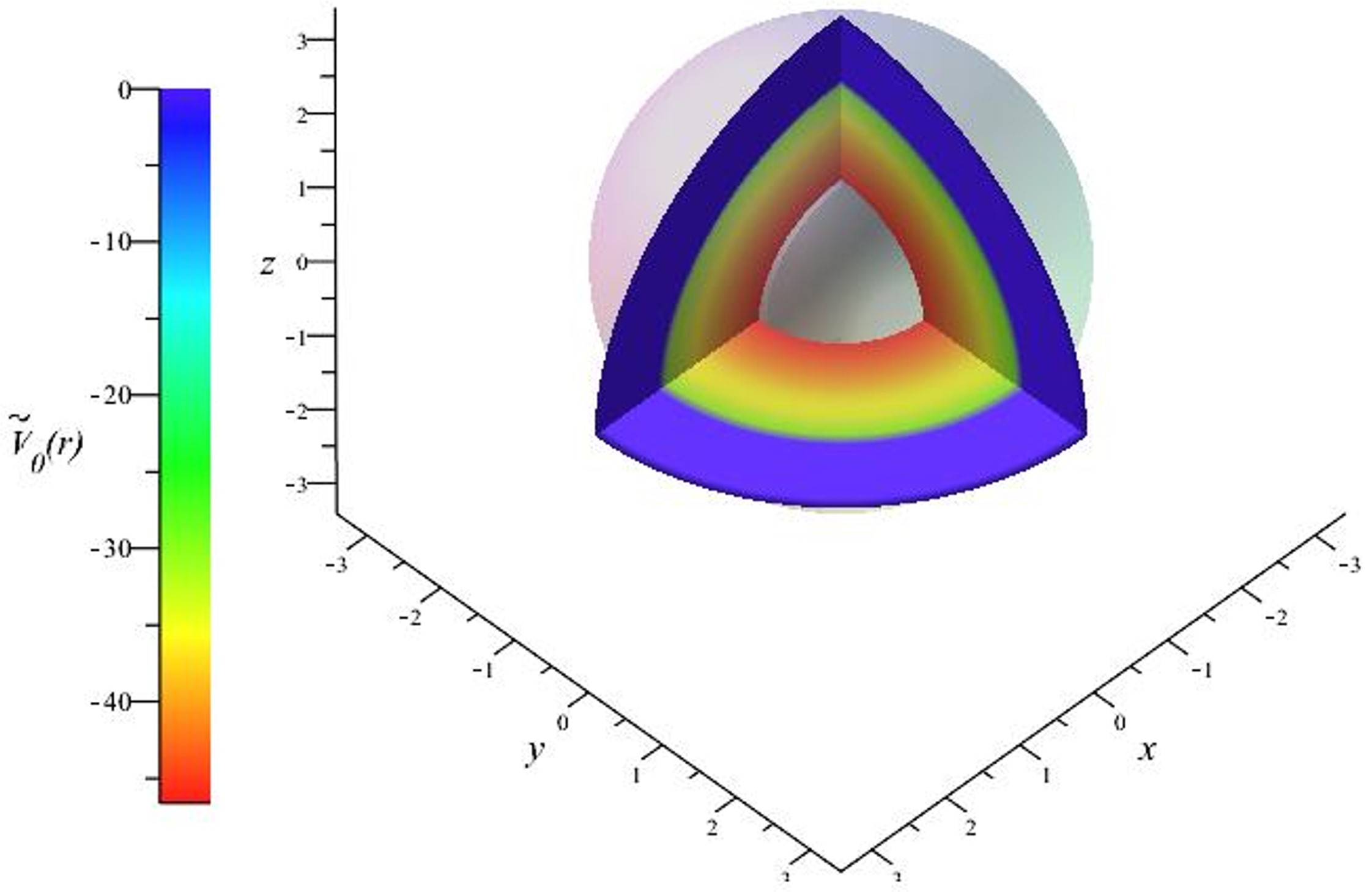}} \ \
    \subfigure[\label{fig:1d}]{\includegraphics[width = .32\linewidth]{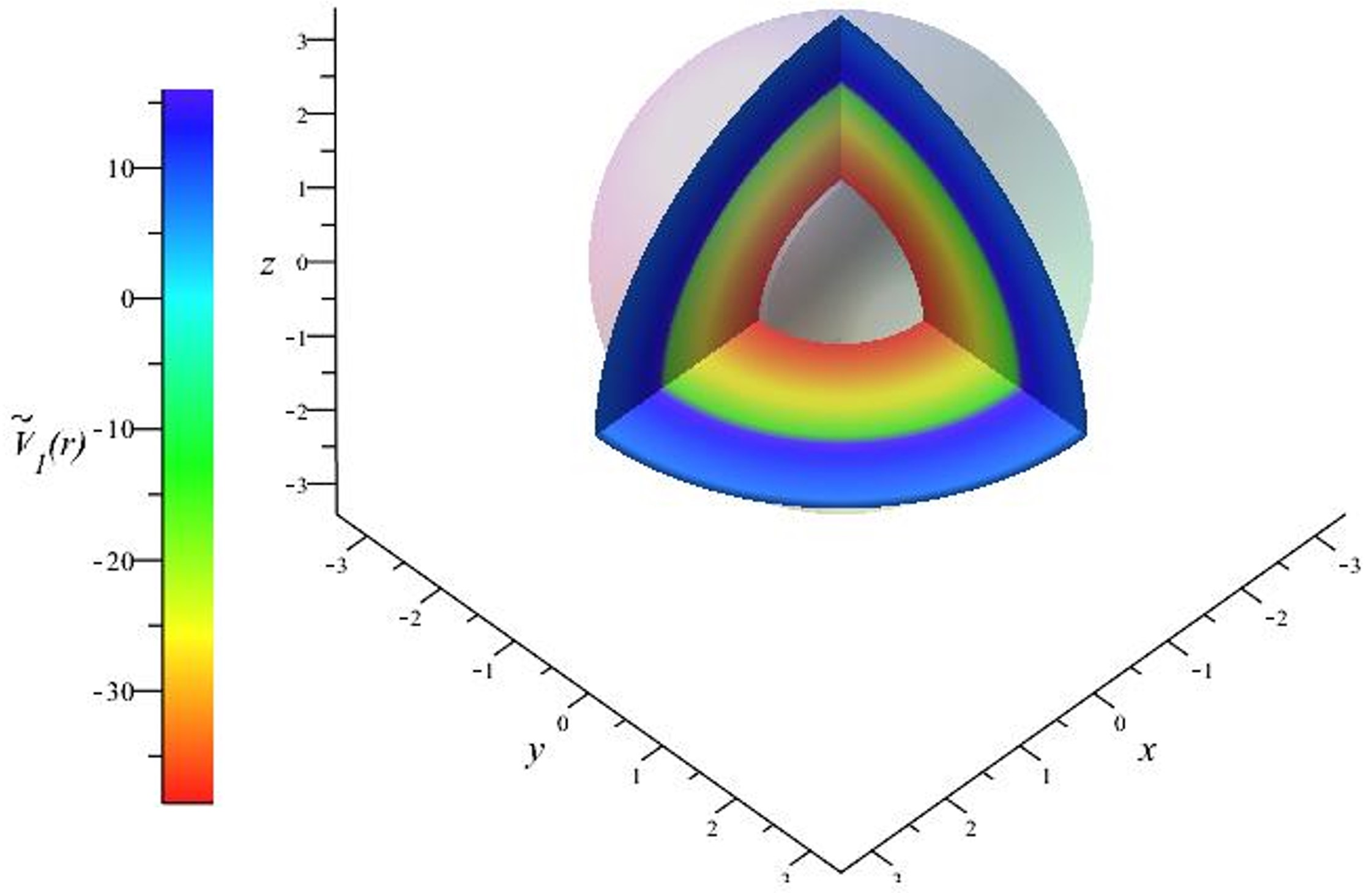}} \ \
    \subfigure[\label{fig:1e}]{\includegraphics[width = .32\linewidth]{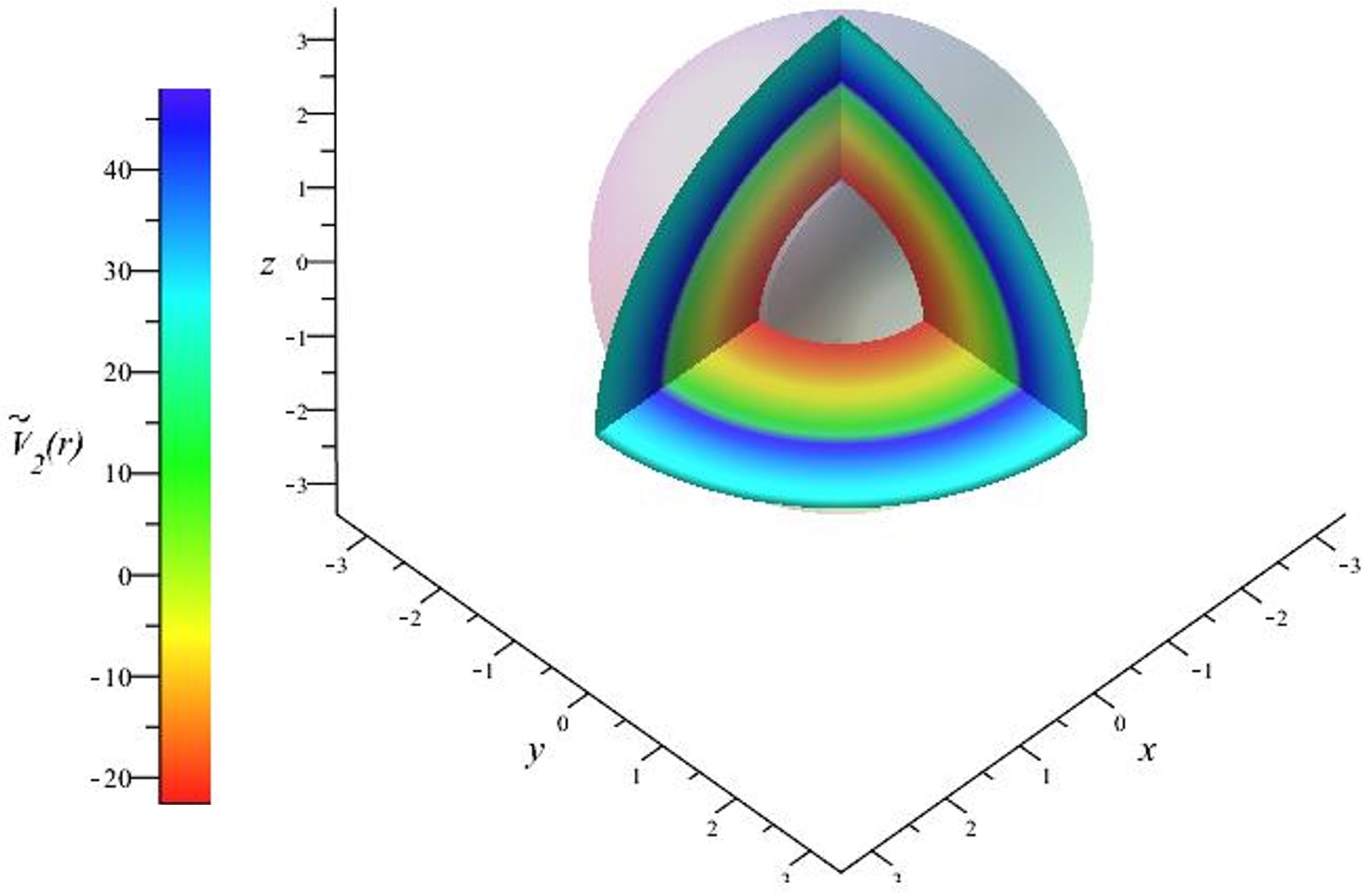}}
    \caption{\subref{fig:1a} Illustration of the model. We show the mass function (magnified $105$ times for a better visualization), the external potential \(\Tilde{V}(r)\) and some effective potentials \(\Tilde{V}_\ell(r)\). Figs.~\subref{fig:1b}, \subref{fig:1c}, \subref{fig:1d} and \subref{fig:1e} allow a 3D depiction of the model by means of a color gradient indicating the radial variation of the functions $m(r)$, \(\Tilde{V}_0(r)\), \(\Tilde{V}_1(r)\) and \(\Tilde{V}_2(r)\), respectively. Numerical values are specified in the vertical color bars. The grey spherical surfaces represent the interfaces at \(r = \delta_1\) and \(r = \delta_2\) (in Fig~\subref{fig:1b} the grey sphere appears covered by a violet one corresponding to the maximal mass value). In these and other plots we adopted \(\delta_1 = 1\), \(\delta_2 = 2.41\) and \(\Tilde{V}_0 = 54.6\).}
    \label{fig:1}
\end{figure}

Following \cite{lima:christiansen:2023b}, we transform variables by means of \(\text{d}{\xi}/{\text{d}x} = m^{1/2}\) and define \(\Xi(\xi)\) such that \(rR(r) = m^{1/4}\Xi(\xi)\) in order to make eq.~\eqref{eq:rgse} a simpler differential equation. In this variables we thus get a Schrodinger regular type equation but in a highly modified potential
\begin{subequations}
    \label{eq:se:mp} % eq: schrödinger equation and modified potential
    \begin{equation}
        \label{eq:se} % subeq: schrödinger equation
        -\Xi''(\xi) + V_{a,b;\ell}(\xi)\Xi(\xi) = \Tilde{E}\Xi(\xi) \; .
    \end{equation}
The total (dimensionless) potential now is
\begin{multline*}
    V_{a,b;\ell}(\xi) \equiv \Tilde{V}(r(\xi)) + \frac{1}{m}\left[\frac{\ell(\ell + 1)}{{r(\xi)}^2} - \frac{1}{2}\left(a + b + \frac{1}{2}\right){\left(\frac{m'}{m}\right)}' + \right. \\
    \left. + \left(ab + \frac{a + b}{2} + \frac{3}{16}\right){\left(\frac{m'}{m}\right)}^2 - \frac{a + b + 1}{r(\xi)}\frac{m'}{m}\right] \; .
\end{multline*}
% COMENT: QUEBREI A EQ. EM UM PONTO DIFERENTE PARA NÃO ULTRAPASSAR A LARGURA DA LINHA

To complete the model and carry on calculations further we must choose a  {PDM} function. We adopt a rapidly decreasing inverse polynomial  \cite{lima:christiansen:2023b,khlevniuk:tymchyshyn:2018} which can be analytically treated
{\[M (\mathscr{r}) =
\begin{cases}
    m_0{\left[1 + {\left(\frac{\mathscr{r}}{\epsilon}\right)}^2\right]}^{-2}    &   \epsilon\delta_1 \leq \mathscr{r} \leq \epsilon\delta_2   \\
    \dfrac{m_0}{{(1 + \delta_2^2)}^2}   &   \mathscr{r} > \epsilon\delta_2
\end{cases}\]}
%(como a partícula não entra no núcleo duro, não precisamos definir a massa para \(\mathscr{r} < \epsilon\delta_1\) \cite{dong:lozada-cassou:2005}).} 
For this mass the effective potential reads
\[\Tilde{V}_\ell(r) =
\begin{cases}
    \infty                                                                      &   r < \delta_1    \\
    -\Tilde{V}_B + \dfrac{\ell(\ell + 1)}{r^2} + [4 + \ell(\ell + 1)](1 + r^2)  &   \delta_1 \leq r \leq \delta_2   \\
    {(1 + \delta_2^2)}^2 \,\dfrac{\ell(\ell + 1)}{r^2}                          &   r > \delta_2
\end{cases}\]
(see Fig.~\ref{fig:1}) and the new variable results \(\xi = \arctan{r}\). In terms of \(\xi\) the {\it total} effective potential can be written as
    \begin{equation}
        \label{eq:mp} % eq: modified potential
        V_{a,b;\ell}(\xi) = \ell(\ell + 1)\sec^2{\xi}\csc^2{\xi} + \omega_{a,b}\tan^2{\xi} + V_{a,b}^{(0)} \; ,
    \end{equation}
\end{subequations}
where \(\omega_{a,b} \equiv 16ab + 10(a + b) + 6\) \, and \, \(V_{a,b}^{(0)} \equiv V_{a,b;0}(0) = 6(a + b) + 5 - \Tilde{V}_B\).
In the internal shell, in contact with the core, the solutions of eq.~\eqref{eq:se} are
\[\Xi_{a,b;\ell}^\pm(\xi) = {\left(\sin{\xi}\right)}^{2\mu_\ell^\pm}{\left(\cos{\xi}\right)}^{2\nu_{a,b;\ell}}\hypergeom{\alpha_{a,b;\ell}^\pm,\beta_{a,b;\ell}^\pm;\gamma_\ell^\pm;\sin^2{\xi}} \; ,\]
where \(\mu_\ell^\pm \equiv \frac{1 \pm (2\ell + 1)}{4}\), \(\nu_{a,b;\ell} \equiv \frac{1}{4} - \frac{1}{2}\sqrt{{\left(\ell + \frac{1}{2}\right)}^2 + \omega_{a,b}}\), \(\alpha_{a,b;\ell}^\pm \equiv \mu_\ell^\pm + \nu_{a,b;\ell} + \frac{1}{2}\sqrt{\Delta_{a,b}}\), \(\beta_{a,b;\ell}^\pm \equiv \mu_\ell^\pm + \nu_{a,b;\ell} - \frac{1}{2}\sqrt{\Delta_{a,b}}\), \(\gamma_\ell^\pm \equiv 2\mu_\ell^\pm + \frac{1}{2}\) e \(\Delta_{a,b} \equiv \Tilde{E} - V_{a,b}^{(0)} + \omega_{a,b}\). But now, we cannot discard \(\Xi_{a,b;\ell}^-(\xi)\) as in \cite{lima:christiansen:2023b} since the origin is outside the domain. Thus, the solutions to be considered are
\begin{equation}
    \label{eq:irps} % eq: internal radial particular solutions
    R_{a,b;\ell}^\pm(r) = \frac{r^{2\mu_\ell^\pm-1}}{{(1 + r^2)}^{\mu_\ell^\pm + \nu_{a,b;\ell} + \frac{1}{2}}}\hypergeom{\alpha_{a,b;\ell}^\pm,\beta_{a,b;\ell}^\pm;\gamma_\ell^\pm;\frac{r^2}{1 + r^2}} \; , \quad \delta_1 \leq r \leq \delta_2 \; .
\end{equation}
The general solution \(R_{a,b;\ell}^\text{in}(r)\) has to fulfill a hard condition due to the core (to vanish at 
 \(r = \delta_1\)) yielding
\begin{equation}
    \label{eq:irgs} % eq: internal radial general solution
    R_{a,b;\ell}^\text{in}(r) = C \left(R_{a,b;\ell}^+(r) - \frac{R_{a,b;\ell}^+(\delta_1)}{R_{a,b;\ell}^-(\delta_1)}R_{a,b;\ell}^-(r)\right) \; ,
\end{equation}
{where \(C \in \mathbb{C}\) is a normalization factor.}

% sec: dinamics in exterior region
\subsection{The external {layer}}
    \label{sec:der}

In the region \(r > \delta_2\), eq.~\eqref{eq:rgse} {reduces to}
\[{(rR_\ell(r))}'' + \left[\frac{\Tilde{E}}{{(1 + \delta_2)}^2} - \frac{\ell(\ell + 1)}{r^2}\right]rR_\ell(r) = 0 \; ,\]
and the general solution is
\begin{equation}
    \label{eq:ergs} % eq: exterior radial general solution
R_\ell^\text{ext}(r) =
\begin{cases}
    Cc^{(1)}R_\ell^{(1)}(r)                             &   \Tilde{E} \leq 0    \\
    Cc^{(1)}R_\ell^{(1)}(r) + Cc^{(2)}R_\ell^{(2)}(r)   &   \Tilde{E} > 0
\end{cases}
\; , \quad c^{(1,2)} \in \mathbb{C} \; ,
\end{equation}
where the linearly independent components are given in terms of the $J$, $K$ and $Y$ Bessel functions:
\begin{equation}
    \label{eq:erps} % eq: exterior radial particular solutions
\begin{cases}
    R_\ell^{(1)}(r) =
    \begin{cases}
        \frac{1}{\sqrt{r}}K_{\ell + \frac{1}{2}}(kr) &   \Tilde{E} < 0  \\
        r^{-(\ell + 1)}                              &   \Tilde{E} = 0  \\
        \frac{1}{\sqrt{r}}J_{\ell + \frac{1}{2}}(kr) &   \Tilde{E} > 0
    \end{cases} \\
    R_\ell^{(2)}(r) = \frac{1}{\sqrt{r}}Y_{\ell + \frac{1}{2}}(kr) \;\;\quad    \Tilde{E} > 0
\end{cases}
\;  ,   \quad   k \equiv \frac{{|\Tilde{E}|}^{1/2}}{(1+\delta_2^2)}    \;  .
\end{equation}
%and $R_\ell^{(2)} = \frac{1}{\sqrt{r}}Y_{\ell + \frac{1}{2}}(kr)$. 
{Here, we have already} discarded the solutions which diverge at infinity.

% sec: boundary conditions
\subsection{More about interface conditions}
    \label{sec:bc}
As we can see in Fig.~\ref{fig:1}, there is a (finite) potential energy gap at \(r = \delta_2\); then we must {discard} orderings GW and LK and keep just $b = a$ \cite{morrow:brownstein:1984}. The solution \(R(r) \equiv R_{a;\ell}(r)\) and its derivative \(\text{d}R_{a;\ell}(r)/\text{d}r\) have to be continuous at \(r = \delta_2\). For this analysis we adopt a vector notation (\(\tau = 1,2\)),
\[\boldsymbol{\varphi}_{a;\ell;\tau} \equiv
\begin{pmatrix}
    R_{a,a;\ell}^+(\delta_\tau) \\
    R_{a,a;\ell}^-(\delta_\tau)
\end{pmatrix}
\; , \; \dot{\boldsymbol{\varphi}}_{a;\ell;\tau} \equiv
\begin{pmatrix}
    \left.\diff{R_{a,a;\ell}^+}{r}\right|_{r = \delta_\tau} \\
    \left.\diff{R_{a,a;\ell}^-}{r}\right|_{r = \delta_\tau}
\end{pmatrix}
\; , \; \boldsymbol{\vartheta}_\ell \equiv
\begin{pmatrix}
    R_\ell^{(1)}(\delta_2) \\
    R_\ell^{(2)}(\delta_2)
\end{pmatrix}
\; \text{and} \; \dot{\boldsymbol{\vartheta}}_\ell \equiv
\begin{pmatrix}
    \left.\diff{R_\ell^{(1)}}{r}\right|_{r = \delta_2} \\
    \left.\diff{R_\ell^{(2)}}{r}\right|_{r = \delta_2}
\end{pmatrix}
\; .\]
% COMENT: AJUSTE NOS ESPAÇAMENTOS ENTRE AS MATRIZES
For a vector \(v^{(a,b)} \in \mathbb{C}\), we also define
\(
{\begin{pmatrix}
    v^{(a)} \\
    v^{(b)}
\end{pmatrix}
}^\perp \equiv
\begin{pmatrix}
    v^{(b)} \\
    -v^{(a)}
\end{pmatrix}
\) and
\(
{\begin{pmatrix}
    v^{(a)} \\
    v^{(b)}
\end{pmatrix}
}^T \equiv
\begin{pmatrix}
    v^{(a)} &   v^{(b)}
\end{pmatrix}
\).

(i) First let us discuss the bound states. For \(\Tilde{E} \leq 0\) the interface conditions can be written as
\begin{equation}
    \label{eq:es} % eq: energy spectrum
    \dot{\boldsymbol{\varphi}}_{a;\mathscr{n},\ell;2}^T\boldsymbol{\varphi}_{a;\mathscr{n},\ell;1}^\perp - \frac{1}{R_{\mathscr{n},\ell}^{(1)}(\delta_2)}\left.\diff{R_{\mathscr{n},\ell}^{(1)}}{r}\right|_{r = \delta_2}\boldsymbol{\varphi}_{a;\mathscr{n},\ell;2}^T\boldsymbol{\varphi}_{a;\mathscr{n},\ell;1}^\perp = 0 \; .
\end{equation}
These result in a discrete energy spectrum \(\Tilde{E} \equiv \Tilde{E}_{a;\mathscr{n},\ell}\) and the following expression for \(c^{(1)} \equiv c_{a;\mathscr{n},\ell}^{(1)}\),
\[c_{a;\mathscr{n},\ell}^{(1)} = \frac{\boldsymbol{\varphi}_{a;\mathscr{n},\ell;2}^T\boldsymbol{\varphi}_{a;\mathscr{n},\ell;1}^\perp}{R_{a,a;\mathscr{n},\ell}^-(\delta_1)R_{\mathscr{n},\ell}^{(1)}(\delta_2)} \; .\]
The new quantum number \(\mathscr{n} \in \mathbb{Z}_+^*\) numerates the \(\mathscr{n}\)th solution of eq.~\eqref{eq:es} and emerges implicitly in the operation. Since the spectrum does not depend on \(\mathscr{m}_\ell\), the \(2\ell + 1\) orbitals \((\mathscr{n},\ell,\mathscr{m}_\ell)\) are degenerated in energy.
Considering this new quantum number, the three-dimensional bound eigenstates are given by \(\psi_{a;\mathscr{n},\ell,\mathscr{m}_\ell}(\boldsymbol{r}) = C_{a;\mathscr{n},\ell}\mathcal{R}_{a;\mathscr{n},\ell}(r)\Upsilon_\ell^{\mathscr{m}_\ell}(\theta,\phi)\) where the coefficients \(C \equiv C_{a;\mathscr{n},\ell}\) are obtained by normalization: \(C_{a;\mathscr{n},\ell} = {\left(\int_{\delta_1}^\infty{\sqabs{\mathcal{R}_{a;\mathscr{n},\ell}(r)}r^2\text{d}r}\right)}^{-1/2}\).

(ii) Now, let us evaluate the continuous spectrum. For \(\Tilde{E} > 0\) the boundary conditions yield the following expressions for \(c^{(1,2)} \equiv c_{a;\ell}^{(1,2)}\):
\begin{equation}
    \label{eq:css} % eq: coefficients scattering states
    \begin{pmatrix}
        c_{a;\ell}^{(1)}    \\
        c_{a;\ell}^{(2)}
    \end{pmatrix}
    = \frac{\boldsymbol{\varphi}_{a;\ell;2}^T\boldsymbol{\varphi}_{a;\ell;1}^\perp}{R_{a,a;\ell}^-(\delta_1)\boldsymbol{\vartheta}_\ell^T\dot{\boldsymbol{\vartheta}}_{\ell}^\perp}\dot{\boldsymbol{\vartheta}}_{\ell}^\perp - \frac{\dot{\boldsymbol{\varphi}}_{a;\ell;2}^T\boldsymbol{\varphi}_{a;\ell;1}^\perp}{R_{a,a;\ell}^-(\delta_1)\boldsymbol{\vartheta}_\ell^T\dot{\boldsymbol{\vartheta}}_{\ell\perp}}\boldsymbol{\vartheta}_{\ell}^\perp \;,
\end{equation}
with no restrictions on the energy values.

\section{Applications and discussion}
    \label{sec:rd}
The new spectrum resulting from {PDM} particles in a multishell structure is useful to model and fit, or predict, the optical features of quantum materials (see e.g. \cite{karabulut:baskoutas:2008, xie:2010, keshavarz:zamani:2013, miranda:mora-ramos:duque:2013, kasapoglu:etal:2021, kasapoglu:etal:2022, yucel:kasapoglu:duque:2022, bayrak:kaya:bayrak:2023, sakiroglu:yucel:kasapoglu:2023}
% \textcolor{blue}{melhor colocar explicitamente as {\bf referências 26, 47 e 71-75 } de \cite{christiansen:lima:2024}}
). For example, the absorption coefficients \(\zeta_{a;i \to f}(E_\gamma)\) and the relative change in the refraction index \({\Delta n}_{a;i \to f}(E_\gamma)/n_r \equiv \delta^{(n)}_{a;i \to f}(E_\gamma)\) of the structure when an incident photon of energy \(E_\gamma\) linearly polarized makes the system transitioning among states \((\mathscr{n}_i,\ell_i,{\mathscr{m}_\ell}_i) \to (\mathscr{n}_f,\ell_f,{\mathscr{m}_\ell}_f)\). In a (first order) perturbative calculation one obtains:
\begin{multline}
    \label{eq:acrric} % eq: absorption coefficient and relative refractive index change
    \zeta_{a;i \to f}(E_\gamma) = \sqrt{\frac{\mu_0}{\varepsilon}}\frac{\kappa_v\Gamma_{i \to f}\sqabs{\mathcal{M}_{a;i \to f}}}{{\left({\Delta E}_{a;i \to f} - E_\gamma\right)}^2 + \Gamma_{i \to f}^2}\frac{E_\gamma}{\hbar} \\
    \text{and} \quad \delta^{(n)}_{a;i \to f}(E_\gamma) = \frac{\kappa_v\sqabs{\mathcal{M}_{a;i \to f}}}{2\varepsilon}\frac{{\Delta E}_{a;i \to f} - E_\gamma}{{\left({\Delta E}_{a;i \to f} - E_\gamma\right)}^2 + \Gamma_{i \to f}^2} \; ,
\end{multline}
where \(\mu_0\) is the permeability of the vacuum, \(\varepsilon = \varepsilon_0n_r^2 = \varepsilon_0\varepsilon_r\) is the real part of the permittivity of the material (\(\varepsilon_0\) is the permittivity of the vacuum, \(\varepsilon_r\) and \(n_r = \sqrt{\varepsilon_r}\) are the static dielectric constant and the refractive index of the medium, respectively), \(\kappa_v\) is the carrier density, \(\Gamma_{i \to f}\) is the relaxation rate, \(\mathcal{M}_{a;i \to f} = e\epsilon \left\langle\left.\psi_{a;\mathscr{n}_f,\ell_f,{\mathscr{m}_\ell}_f}\right|z\left|\psi_{a;\mathscr{n}_i,\ell_i,{\mathscr{m}_\ell}_i}\right.\right\rangle\) is the dipole matrix element for the $z$-polarized incident radiation ($e$ is the elementary charge) and \({\Delta E}_{a;i \to f} \equiv E_{a;\mathscr{n}_f,\ell_f,{\mathscr{m}_\ell}_f} - E_{a;\mathscr{n}_i,\ell_i,{\mathscr{m}_\ell}_i}\) is the energy gap between the levels \cite{kasapoglu:etal:2022}. {Note that the internal product is invariant \(\left\langle\left.\psi_1\right|f\left|\psi_2\right.\right\rangle=\left\langle\left.\Psi_1\right|f\left|\Psi_2\right.\right\rangle \) under the transformations performed along the calculations.}

\begin{figure}
    \centering
    \includegraphics[width=0.5\linewidth]{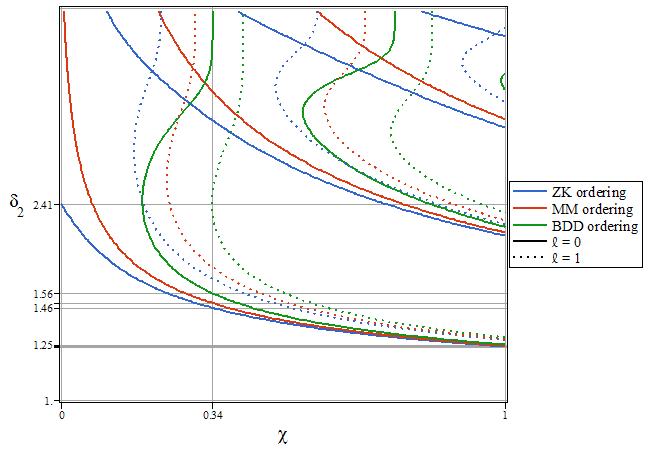} 
    %ATENCION: ELSEVIER NO COMPILA poniendo en includegraphics {Figures/fig} - Hay que poner apenas {fig} 
    \caption{{Graphics of  \(\delta_2\) vs. \(\chi\) in the imminence of freedom. For a given value of the variables,  \((\delta_2^0, \chi^0)\), we can see the bound states allowed just watching the vertical line below \(\delta_2^0\).}}
    \label{fig:wc}
\end{figure}

In order to apply our results to phenomenological situations, we have to fix the parameters. If we consider a heterostructure like \(\text{GaAs}/\text{A}\ell\text{GaAs}\) we may adopt $m_0 \equiv m_\text{GaAs} = 0.067m_e$ ($m_e$ being the free electron mass) and \(\varepsilon \equiv \varepsilon_\text{GaAs} = 13.18\varepsilon_0\). Taking \(\delta_1 = 1\) and \(\epsilon\) as the effective Bohr radius of \(\text{GaAs}\) we get \(\epsilon = 4\pi\varepsilon_\text{GaAs}\hbar^2/m_\text{GaAs}e^2 = \qty{10.4}{\nm}\); with this, the energy conversion factor \(\mathcal{E}\) is given by the Rydberg effective constant, \(\mathcal{E} = m_\text{GaAs}e^4/8\varepsilon_\text{GaAs}^2h^2 = \qty{5.25}{\meV}\) \cite{naimi:vahedi:soltani:2015}. 
{The depth of the potential well is related to the concentration \(\chi\) of Aluminium in \({\text{A}\ell}_\chi\text{Ga}_{1-\chi}\text{As}\) through \(\mathcal{E}\Tilde{V}_B(\chi) = 658(1.155\chi + 0.370\chi^2)\;\unit{\meV}\) \cite{kasapoglu:etal:2022}. Both, \(\chi\) and \(\delta_2\),  can be adjusted in order to obtain at least two bound states in the three von Roos orderings so as to allow a comparison of the effect of operator's ordering in the optical properties of the system. With this in mind, in Fig.~\ref{fig:wc} we depicted 
eq.~\eqref{eq:es} by means of \(\delta_2\) versus \(\chi\) for \(\Tilde{E} = 0\) (i.e., we evaluate the points \((\chi,\delta_2)\) in the imminence of the continuum). Through these calculations we verified that the six bound states take place for \(\chi \geq 0.34\),  the critical values being \(\chi = 0.34\), \(\delta_2 = 2.41\). With this critical concentration, the potential depth results \(\Tilde{V}_B = 54.6\).}
%Here we adopt the smallest concentration \(\chi\) for which all the von Roos orderings have at least two bound states for some \(\delta_2\), to allow a comparison of the effect of ordering in the optical properties of the system. After inspection, we {set} \(\chi \approx 0.340\) which implies \(\Tilde{V}_B = 54.6\) and \(\delta_2 = 2.41\).

\begin{figure}
    \subfigure[\label{fig:2a}]{\includegraphics[width = .49\linewidth]{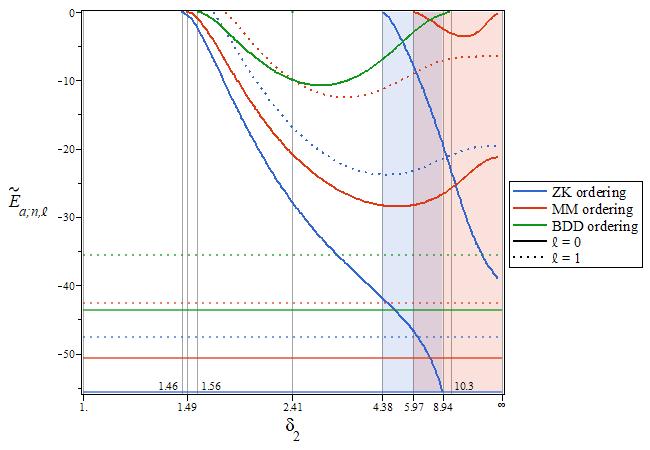}} \ \
    \subfigure[\label{fig:2b}]{\includegraphics[width = .49\linewidth]{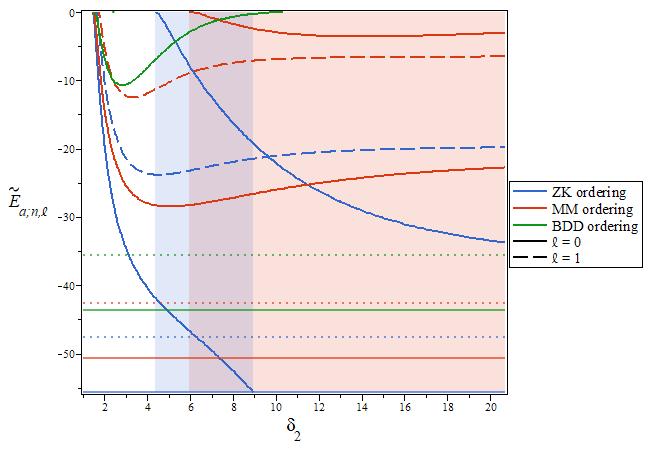}}
    \caption{{Dependence of the energy eigenvalues with the external wall position \(\delta_2\) of the shell  for \(\chi = 0.34\). The horizontal lines represent the effective potential minima (independent of \(\delta_2\)). In \subref{fig:2a} the curves are deformed as a consequence of scale modification to allow arbitrary large values of the second zone; in \subref{fig:2b} we restricted the interval of \(\delta_2\) to show the exact shape of the curves.}}
    \label{fig:2}
\end{figure}

Before completing this discussion, we open a parenthesis about the variation of the bounding energies with the width of the inner shell. {In Fig.~\ref{fig:2} we plot these eigenenergies with respect to the interface radius \(\delta_2\) for the BDD, ZK and MM orderings, keeping \({\text{A}\ell}\) concentration at \(\chi = 0.34\). We can see that the shell's external wall corresponding to \(\Tilde{E} = 0\) (top line)  matches the crossing points of the \(\delta_2\) lines  with the vertical line \(\chi = 0.34\) in Fig.~\ref{fig:wc}}.
{Some comments are in order. In Fig.~\ref{fig:2} we note that the fundamental states of every ordering (solid lines) start at a minimal value of \(\delta_2\), which depends on the ordering. Specifically, for  ZK, MM and BDD the values are respectively \(\delta_{2,\text{min}} = 1.46\), $1.49$ and $1.56$ (see solid horizontal lines in Fig.~\ref{fig:wc}). It imposes a minimum shell-width, for each ordering, for the existence of bound states which depends on the concentration of \({\text{A}\ell}\). The smallest width are obtained as  \(\chi \to 1\), for which  \(\delta_2 \to 1.24\) in the  ZK and MM orderings and \(\delta_2 \to 1.25\) for BDD.}
%In Fig.~\ref{fig:2} note the beginning of the bound energy curves on the upper left. It means that for \(\delta_2 < 1.46\delta_1\) there are no bound states. It thus establishes a minimum shell width for a given core radius.
{Indeed, the variation of  \(\Tilde{V}_B\) with \(\chi\) indicates a growing depth in the total effective potential (eq.~\eqref{eq:mp}) with the increasing concentration which  will allow smaller minimal shells.
The particular value of each minimal shell-width is naturally a result of the solutions of the modified differential equation, which is determined by the mass and the different kinetic potential shapes. The depth of the effective potential is a phenomenological function of the   \({\text{A}\ell}\) concentration shown above, which determines a higher or a lower number of alowed bound-states. In Fig.~\ref{fig:wc} we showed the minimal thickness for a concentration \(\chi\)=0.34. For other concentrations the differential equation, and its solutions, will change and consequently will the minimal shell-width. }

Following with the analysis of Fig.~\ref{fig:2}, there are two solid blue energy curves corresponding to the ordering ZK with \(\ell = 0\). In the colored region \(4.38 < \delta_2 < 8.94\) we identify both the  {\((\mathscr{n},\ell) = (1,0)\)} line below and  {\((\mathscr{n},\ell) = (2,0)\)} above. In these cases the bounding energy is monotonically decreasing with \(\delta_2\). The solid red energy lines correspond to the ordering MM with \(\ell = 0\). They exist together for \(\delta_2 > 5.97\) and again the line below is the bounding energy for quantum numbers {\((\mathscr{n},\ell) = (1,0)\)} and above for  {\((\mathscr{n},\ell) = (2,0)\)}. These case curves are convex, have a minimum value at some \(\delta_2\) and then grow up to zero as \(\delta_2 \to \infty\). In the case of the BDD ordering, there is only one energy curve (for this concentration) which is also convex and quite limited in the width of the inner shell (\(\delta_{2 {\text{max}}} = 10.3\)). We have admitted here a zero energy value as a limiting bound state taking place at \(\delta_2 = 2.41\).

\begin{figure}
    \centering
    \includegraphics[width = .5\linewidth]{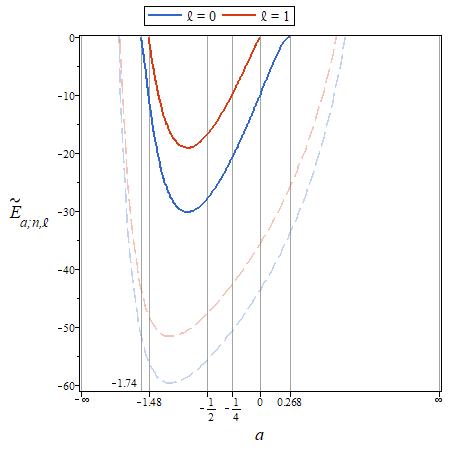}
    \caption{{Bound eigenergies as a function of $a$. The dashed lines depict the minimum value of the total potential \({V_{a;\ell}}_\text{min}\), which is obtained by setting \(\xi = \frac{\pi}{4}\) and $b = a$ in eq.~\eqref{eq:mp}, resulting in a quadratic dependence on $a$ (this shape is deformed due to the change of scale adopted to cover the whole real interval).}}
    \label{fig:3}
\end{figure}

Back to the selected values \(\delta_2/\delta_1 = 2.41\) and \(\Tilde{V}_B = 54.6\), we obtain the eigenstate energies, \( \Tilde{E}_{a;\mathscr{n},\ell}\), for the ZK ( {$a = -1/2$}), MM ( {$a = -1/4$}) and BDD ($a = 0$) orderings. The fundamental energy values are respectively \(\Tilde{E}_{-\frac{1}{2};1,0} = -27.9\), \(\Tilde{E}_{-\frac{1}{4};1,0} = -20.8\) and \(\Tilde{E}_{0;1,0} = -9.97\). In the case of the first excited states, the results are also quite different depending on the kinetic Hamiltonian: \(\Tilde{E}_{-\frac{1}{2};1,0} = -16.8\), \(\Tilde{E}_{-\frac{1}{4};1,0} = -9.90\) and \(\Tilde{E}_{0;1,0} = 0\). In Fig.~\ref{fig:3}, we let the ordering parameter $a$ vary freely to see the values that allow up to two bound states. We find the interval \(-1.74 < a < 0.268\) for the fundamental state value \(\ell = 0\)  {(blue shaded region)} and {\(-1.4828 < a < 0\)} for both \(\ell = 0\) and \(\ell = 1\)  {(red shaded region)}.
\begin{figure}
    \subfigure[\label{fig:4a}]{\includegraphics[width = .49\linewidth]{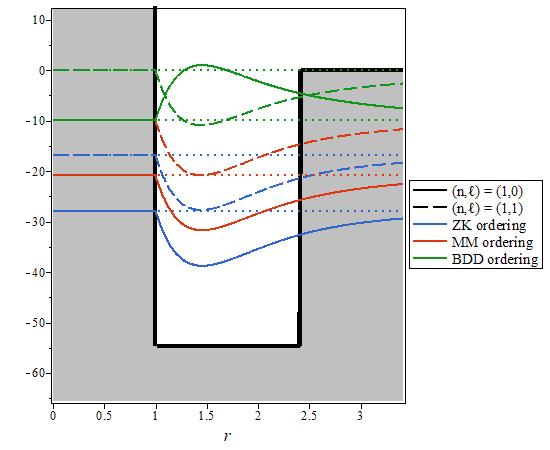}} \ \
    \subfigure[\label{fig:4b}]{\includegraphics[width = .39\linewidth]{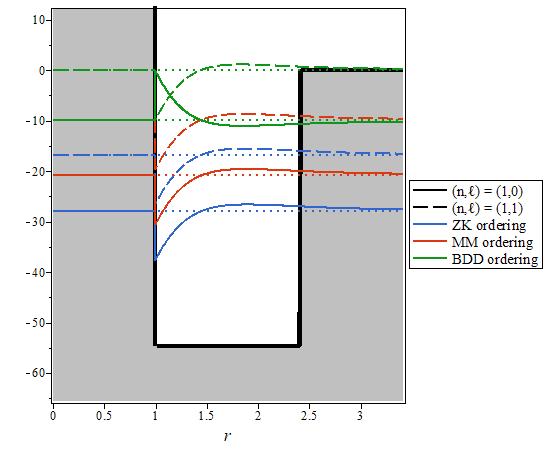}}
    \caption{\subref{fig:4a}  Bound state radial wave functions for the three orderings; \subref{fig:4b} their first derivatives. Here the width  \((\delta_2-\delta_1)/\delta_1=2.41\) and \(\Tilde{V}_B=54.6\)}
    \label{fig:4}
\end{figure}
Next, in Fig.~\ref{fig:4} we show the bound state radial wave functions for the three orderings. We can see that the BDD fundamental state is concave while ZK and MM are convex. The BDD first excited state flips its concavity but the ZK and MM first excited states do not. The continuity of these wave functions and their derivatives can been ascertained.

The bound eigenstate information is then used to calculate the absorption coefficients and the relative change of the refraction index (eq.~\eqref{eq:acrric}) of the system with the above data, for which \(\kappa_v = \qty{3.0e22}{\per\metre\cubed}\), \(\Gamma_{i \to f} = \hbar/(\qty{1}{\ps}) = \qty{0.66}{\meV}\) \cite{kasapoglu:etal:2022}. Dipole selection rules \cite[eq.~17.2.21, p.~459]{shankar:2013} indicate that \(\mathcal{M}_{a;i \to f}\) vanish except for \(\Delta\ell = \pm1\) and \(\Delta\mathscr{m}_\ell = 0\); then it is only allowed the single transition \((1,0,0) \to (1,1,0)\). 

\begin{figure}
    \subfigure[\label{fig:5a}]{\includegraphics[width = .49\linewidth]{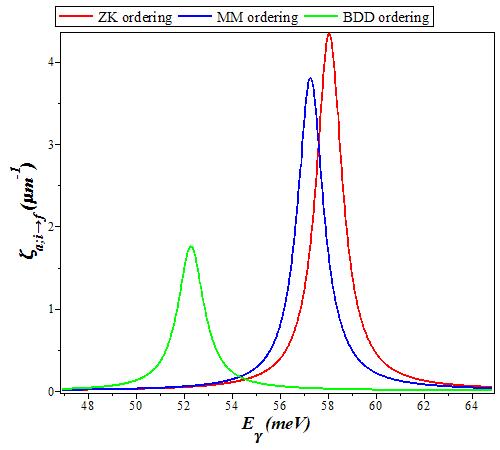}} \ \
    \subfigure[\label{fig:5b}]{\includegraphics[width = .49\linewidth]{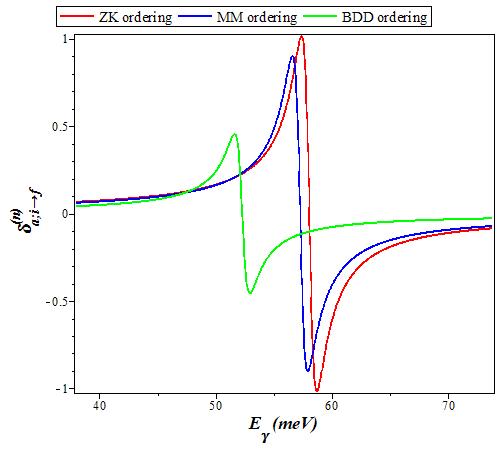}}
    \caption{\subref{fig:5a} Absorption coefficients and \subref{fig:5b} relative change of the refraction index, for transition \((1,0,0) \to (1,1,0)\).}
    \label{fig:5}
\end{figure}

In Fig.~\ref{fig:5} we show the results as a function of the incident photon energy. We can see that the maximal values of the absorption coefficients grow depending on the ordering \(\text{BDD} \to \text{MM} \to \text{ZK}\) and occur at increasing values of the photon energy, accordingly, see Fig.~\ref{fig:5a}. Exactly the same happens with the relative change of the refraction index as is clearly seen in Fig.~\ref{fig:5b}. We can also observe that for low and high photon energies % \(E_\gamma \approx 0\) ou \(E_\gamma \gg 1\) 
the values of both quantities are coincident for any ordering, BDD converging more rapidly to zero.
\vskip 0.5cm

{These optical properties are proportional to the dipole matrix elements and simple rational functions of the transition energy in the first order perturbative calculation shown in eq.~\eqref{eq:acrric}. We can see in 
Fig.~\ref{fig:3} that after a certain value of the ordering parameter $a$, the effective-potential minimum \({V_{a;\ell}}_\text{min}\) is strictly growing for any orbital number \(\ell\); in particular, for the ZK, MM and BDD orderings $a= -0.5 , -0.25$,  and 0,0. Thus, the effective potential deformation originated from the  $a$ dependence of the kinetic potential, allows decreasingly negative eigenvalues with smaller gaps between them as $a$ grows. 
At the same time, the corresponding eigenfunctions vary so that dipole elements follow the same trend and they stregthten together in the variation of the optical properties with the ordering parameter; e.g. \( \Delta E_{ZK, MM, BDD}=11.063 ,  10.914 ,  9.965 ,\) and 
\( \sqabs{\mathcal{M}_{ZK, MM, BDD}}/e\epsilon=0.5986 ,  0.5320 ,  0.2690 \).
This variation is consistent with the blue shift of the absorption coefficients and the relative change of the refraction index in the sequence \(\text{BDD} \to \text{MM} \to \text{ZK}\) shown in Fig.~\ref{fig:3}.}

 {Now, let us assume an incident beam of plane waves \(\Psi_\text{inc}(\boldsymbol{{\mathscr{r}}}) =  \epsilon^{-3/2}\psi_\text{inc}(\boldsymbol{r})\). Assuming it is coming along the $z$ axis,  \(\Psi_\text{inc}(\boldsymbol{{\mathscr{r}}}) = \mathcal{I}e^{ikz}\). It is then spherically scattered by the quantum structure so that  \(\Psi_\text{sc}(\boldsymbol{ {\mathscr{r}}}) = \epsilon^{-3/2}\psi_\text{sc}(\boldsymbol{r}) = \mathcal{I}f(\theta)\frac{e^{ikr}}{\epsilon r}\). In this case, the resulting state \(\Psi_E(\boldsymbol{ {\mathscr{r}}}) = \Psi_\text{inc}(\boldsymbol{ {\mathscr{r}}}) + \Psi_\text{sc}(\boldsymbol{ {\mathscr{r}}})\) is independent of the azimuthal angle \(\phi\) and thus we can represent it in terms of the subset of eigenstates \(\Psi_{a;\ell,0}(\boldsymbol{ {\mathscr{r}}})\) corresponding to \(\mathscr{m}_\ell = 0\), \cite[Cap.~19, p.~531]{shankar:2013}). For a finite range potential, at large distances the radial wave function behaves like a free wave but with a phase shift defined in terms of the specific coefficients determined with eq.~\eqref{eq:css}: \(\varsigma_{a,\ell} = \tan^{-1}({c_{a;\ell}^{(2)}}/{c_{a;\ell}^{(1)}})\). The (dimensional) scattering amplitude results
\[f_a(\theta) = \frac{\epsilon}{k}\sum_{\ell = 0}^\infty{(2\ell + 1)e^{i\varsigma_{a,\ell}\ell}}\sin{\varsigma_{a,\ell}}P_\ell(\cos{\theta})\]
(where \(P_\ell(\omega)\) are the Legendre polynomials) and so the total cross section for each von Roos ordering reads
\begin{equation}
    \label{eq:tss} % eq: total scattering section
    \sigma_a = \int{\sqabs{f_a(\theta)}\text{d}\Omega} = \frac{4\pi\epsilon^2}{k^2}\sum_{\ell = 0}^\infty{(2\ell + 1)\sqabs{\sin{\varsigma_{a,\ell}}}} \; .
\end{equation}
The complete solution is \(\psi_{a;\ell,\mathscr{m}_\ell}(\boldsymbol{r}) = C_{a;\ell}\mathcal{R}_{a;\ell}(r)\Upsilon_\ell^{\mathscr{m}_\ell}(\theta,\phi)\), and \(C_{a;\ell}\) can be expressed in terms of \(\mathcal{I}\),
\[C_{a;\ell} = \mathcal{I}\sqrt{\frac{\pi}{2k}}\frac{e^{i\pi\ell/2}}{c_{a;\ell}^{(1)}-c_{a;\ell}^{(2)}}\]
which is experimentally adjustable.}

\begin{figure}
    \centering
    \includegraphics[width = \linewidth]{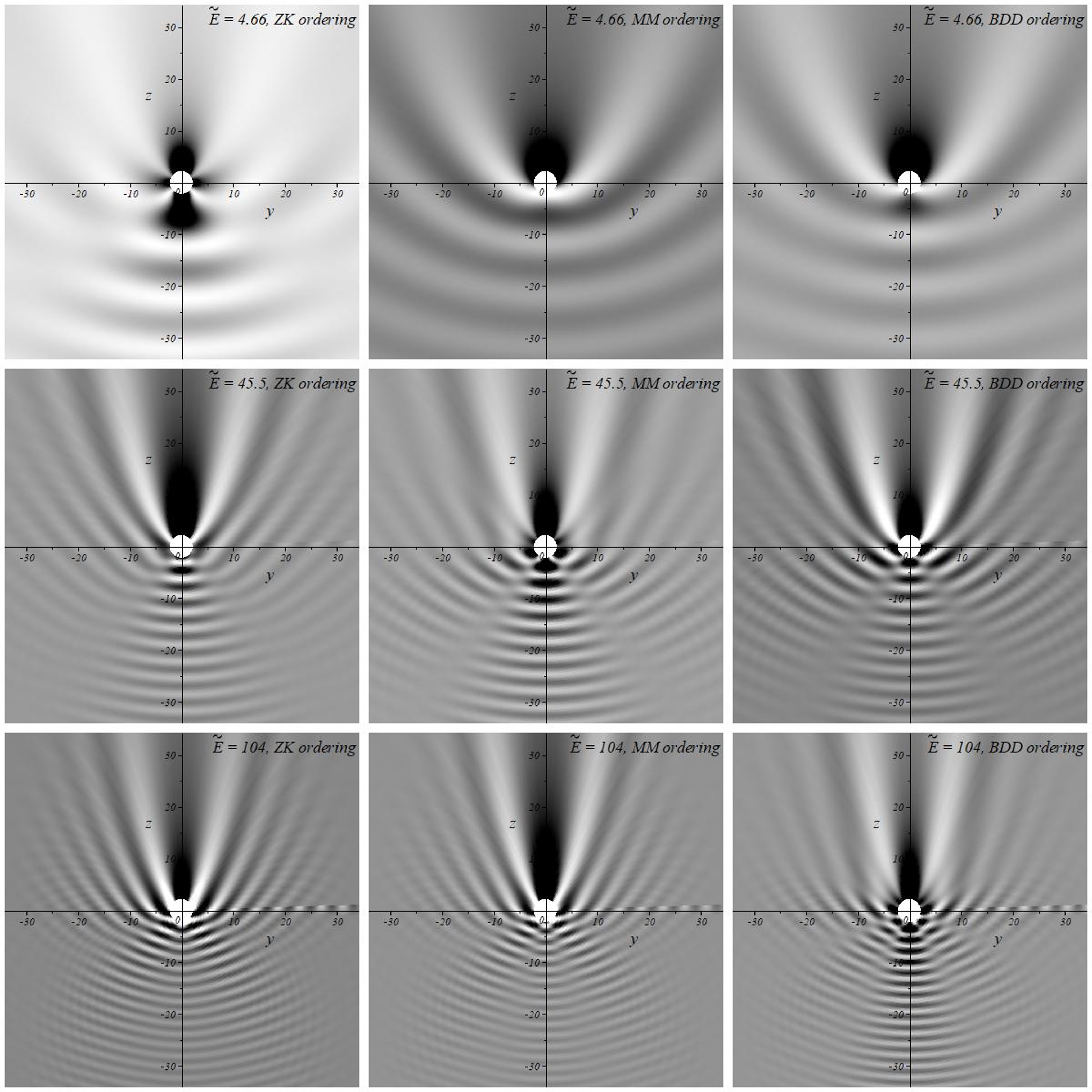}
    \caption{Probability densities \(\sqabs{\psi_{a,\Tilde{E}}}\) for the three orderings ZK, MM e BDD (left to right respectively) for energies \(\Tilde{E} = 4.66\), \(\Tilde{E} = 45.5\) and \(\Tilde{E} = 104\). We chose these energy values to cover a wide range of interest. These correspond, respectively, to the
    first peak in the cross section \(\sigma_a\) in the ZK ordering, to the second in MM and third in BDD (up to down). The central white disk indicates the whole interaction zone where the external potential is nonzero, \(r \leq \delta_2\).}
    \label{fig:6}
\end{figure}

%Regarding the scattering eigenstates, in 
 {In} Fig.~\ref{fig:6} we show the probability densities \(\sqabs{\psi_{a,\Tilde{E}}}\) for the three allowed orderings and three energy levels $\Tilde{E}$. We can see the interference pattern between the incident wave \(\psi_\text{inc}(r) \propto e^{ikz}\) (coming downward from \(z \to \infty\)) and the scattered wave \(\psi_\text{sc}(r) \propto f_a(\theta)\frac{e^{ikr}}{r}\). The darker zones indicate constructive interference while the brighter show destructive interference. We can use these eigenstates for the calculation of the cross sections of the system.

\begin{figure}
    \subfigure[\label{fig:7a} ]{\includegraphics[width = .49\linewidth]{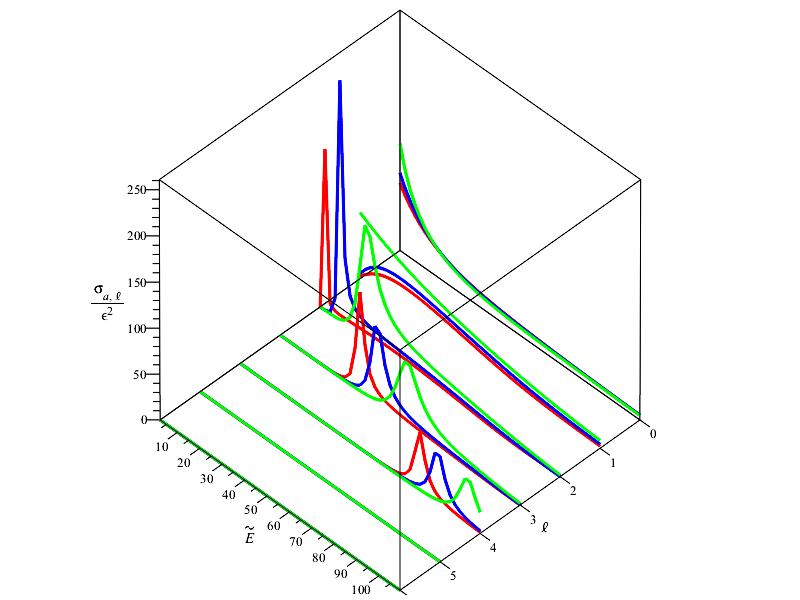}} \ 
    \subfigure[\label{fig:7b} ]{\includegraphics[width = .49\linewidth]{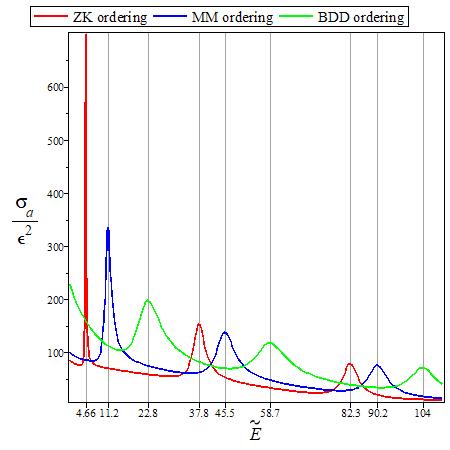}}
    \caption{\subref{fig:7a} Partial scattering cross sections \(\sigma_{a,\ell}\) and \subref{fig:7b} total scattering cross sections \(\sigma_a\), for ZK, MM and BDD orderings (in red, blue and green, respectively).}
    \label{fig:7}
\end{figure}

We plot in Fig.~\ref{fig:7} the partial \(\sigma_{a,\ell}\) and total \(\sigma_a\) cross sections for each different ordering, see eq.~\eqref{eq:tss}. % for \(\delta_1\), \(\delta_2\) e \(\Tilde{V}_B\) usados para os estados ligados e para faixas de energia inferiores a \(2\Tilde{V}_B = 109\). 
A clean maximum is seen in each case. The partial wave maxima occur at higher energies as the higher is the orbital number. At the same time the peaks get lower, see Fig.~\ref{fig:7a}. For \(\ell > 6\) the peaks are negligible as compared to the previous ones. For the total cross section we again note that the sharpest and highest curve corresponds to the ZK ordering, followed by the MM and BDD choices, as in the previous analysis (this rule does not hold for the lower partial cross sections, see Fig.~\ref{fig:7a}). The exact energy values are shown in the plot, see Fig.~\ref{fig:7b}.

\newpage

% sec: Conclusions
\section{Conclusions}
    \label{sec:c}

In this work we have solved the eigenvalue problem of an effective particle with a continuous position-dependent  mass in a multishell nanostructure.  The heterostructure consists of a finite hard core with a potential well crust where the mass varies radially, and a third region where the potential is higher and the mass remains constant. 
It  may represent a mixed semiconductor compound surrounding an isolant, of interest in optoelectronics, or even nuclear matter systems. The analytical treatment of the differential equation and boundary conditions leads to nontrivial radial solutions (see eqs.~\eqref{eq:irps}, \eqref{eq:irgs} and eqs.~\eqref{eq:ergs}, \eqref{eq:erps}) which we found together with the complete eigenvalue spectrum. We have discussed applications of the discrete bound eigenstate sector to the calculation of optical features of quantum heterostructures such as absorption coefficients and refraction indexes. We have shown their dependence with the model parameters such as the size of the regions, the depth of the potential-well and, more importantly, with the mass function. In the continuum eigenstate sector we have analysed the partial and total scattering cross sections of the system.
For both the discrete and continuum spectra we have performed all these calculations for the most relevant Hamiltonian orderings and compared the different outcomes.

\section{Acknowledgement}

The authors would like to thank Conselho Nacional de Desenvolvimento Científico e Tecnológico (CNPq) for partial support.
% Fundo Nacional de Desenvolvimento da Educação do Ministério da Educação (FNDE) for a Programa de Educação Tutorial fellowship.
\\

% bibliography
\bibliographystyle{ieeetr}
\bibliography{refs}

\begin{thebibliography}{10}

\bibitem{wannier:1937}
G.~H. Wannier, ``The structure of electronic excitation levels in insulating crystals,'' {\em Phys. Rev.}, vol.~52, no.~3, p.~191, 1937.

\bibitem{slater:1949}
J.~C. Slater, ``Electrons in perturbed periodic lattices,'' {\em Phys. Rev.}, vol.~76, no.~11, p.~1592, 1949.

\bibitem{luttinger:kohn:1955}
J.~M. Luttinger and W.~Kohn, ``Motion of electrons and holes in perturbed periodic fields,'' {\em Phys. Rev.}, vol.~97, no.~4, p.~869, 1955.

\bibitem{bendaniel:duke:1966}
D.~J. BenDaniel and C.~B. Duke, ``Space-charge effects on electron tunneling,'' {\em Phys. Rev.}, vol.~152, no.~2, p.~683, 1966.

\bibitem{gora:williams:1969}
T.~Gora and F.~Williams, ``Theory of electronic states and transport in graded mixed semiconductors,'' {\em Phys. Rev.}, vol.~177, no.~3, p.~1179, 1969.

\bibitem{vonroos:1983}
O.~von Roos, ``Position-dependent effective masses in semiconductor theory,'' {\em Phys. Rev. B}, vol.~27, no.~12, p.~7547, 1983.

\bibitem{bastard:1992}
G.~Bastard, {\em Wave mechanics applied to semiconductor heterostructures}.
\newblock Les Editions de Physique, 1992.

\bibitem{whalley:etal:2019}
L.~D. Whalley, J.~M. Frost, B.~J. Morgan, and A.~Walsh, ``Impact of nonparabolic electronic band structure on the optical and transport properties of photovoltaic materials,'' {\em Physical review B}, vol.~99, no.~8, p.~085207, 2019.

\bibitem{zhao:liang:ban:2003}
F.~Zhao, X.~Liang, and S.~Ban, ``Influence of the spatially dependent effective mass on bound polarons in finite parabolic quantum wells,'' {\em The European Physical Journal B-Condensed Matter and Complex Systems}, vol.~33, pp.~3--8, 2003.

\bibitem{christiansen:lima:2024}
H.~R. Christiansen and R.~M. Lima, ``Three-dimensional bound states of cylindrical quantum heterostructures with position-dependent mass carriers,'' {\em Physica Scripta}, vol.~99, no.~015915, 2024.

\bibitem{saperstein:etal:2016}
E.~Saperstein, M.~Baldo, N.~Gnezdilov, and S.~Tolokonnikov, ``Phonon effects on the double mass differences in magic nuclei,'' {\em Physical Review C}, vol.~93, no.~3, p.~034302, 2016.

\bibitem{cunha:christiansen:2013}
M.~S. Cunha and H.~R. Christiansen, ``{Analytic results in the position-dependent mass Schr\"odinger problem},'' {\em Commun. Theor. Phys.}, vol.~60, no.~6, p.~642, 2013.

\bibitem{christiansen:cunha:2013}
H.~R. Christiansen and M.~S. Cunha, ``Solutions to position-dependent mass quantum mechanics for a new class of hyperbolic potentials,'' {\em J. Math. Phys.}, vol.~54, no.~12, p.~122108, 2013.

\bibitem{christiansen:cunha:2014}
H.~R. Christiansen and M.~S. Cunha, ``Energy eigenfunctions for position-dependent mass particles in a new class of molecular {H}amiltonians,'' {\em J. Math. Phys.}, vol.~55, no.~9, p.~092102, 2014.

\bibitem{lima:christiansen:2023a}
R.~M. Lima and H.~R. Christiansen, ``The kinetic {H}amiltonian with position-dependent mass,'' {\em Physica E: Low-dimensional Systems and Nanostructures}, vol.~150, no.~115688, 2023.

\bibitem{ho:roy:2019}
C.-L. Ho and P.~Roy, ``Generalized {D}irac oscillators with position-dependent mass,'' {\em EPL-Europhys. Lett.}, vol.~124, no.~6, p.~60003, 2019.

\bibitem{schmidt:dejesus:2018}
A.~G.~M. Schmidt and A.~L. de~Jesus, ``Mapping between charge-monopole and position-dependent mass systems,'' {\em J. Math. Phys.}, vol.~59, no.~10, p.~102101, 2018.

\bibitem{lima:christiansen:2023b}
R.~M. Lima and H.~R. Christiansen, ``Energy eigenstates of position-dependent mass particles in a spherical quantum dot,'' {\em The European Physical Journal B}, vol.~96, no.~150, 2023.

\bibitem{sari:etal:2019}
H.~Sari, E.~Kasapoglu, S.~Sakiroglu, I.~S{\"o}kmen, and C.~A. Duque, ``{Effect of Intense Laser Field in Gaussian Quantum Well With Position-Dependent Effective Mass},'' {\em Physica Status Solidi B}, vol.~256, no.~8, p.~1800758, 2019.

\bibitem{elnabulsi:2020}
R.~A. El-Nabulsi, ``A generalized self-consistent approach to study position-dependent mass in semiconductors organic heterostructures and crystalline impure materials,'' {\em Phys. E: Low Dim. Syst. Nanostruct.}, vol.~124, p.~114295, 2020.

\bibitem{elnabulsi:2020b}
R.~A. El-Nabulsi, ``A new approach to schrodinger equation with position-dependent mass and its implications in quantum dots and semiconductors,'' {\em J. Phys. Chem. Sol.}, vol.~140, p.~109384, 2020.

\bibitem{kasapoglu:etal:2021}
E.~Kasapoglu, H.~Sari, I.~S{\"o}kmen, J.~A. Vinasco, D.~Laroze, and C.~A. Duque, ``Effects of intense laser field and position dependent effective mass in {R}azavy quantum wells and quantum dots,'' {\em Physica E: Low-dimensional Systems and Nanostructures}, vol.~126, p.~114461, 2021.

\bibitem{kasapoglu:duque:2021}
E.~Kasapoglu and C.~A. Duque, ``Position dependent effective mass effect on the quantum wells with three-parameter modified manning potential,'' {\em Optik}, vol.~243, p.~166840, 2021.

\bibitem{valencia-torres:etal:2020}
R.~Valencia-Torres, J.~Avenda{\~n}o, A.~Bernal, and J.~Garc{\'\i}a-Ravelo, ``Energy spectra of position-dependent masses in double heterostructures,'' {\em Physica Scripta}, vol.~95, no.~7, p.~075207, 2020.

\bibitem{ganguly:etal:2006}
A.~Ganguly, {\c{S}}.~Kuru, J.~Negro, and L.~Nieto, ``A study of the bound states for square potential wells with position-dependent mass,'' {\em Physics Letters A}, vol.~360, no.~2, pp.~228--233, 2006.

\bibitem{galbraith:duggan:1988}
I.~Galbraith and G.~Duggan, ``{Envelope-function matching conditions for GaAs/(Al,Ga)As heterojunctions},'' {\em Phys. Rev. B}, vol.~38, no.~14, p.~10057, 1988.

\bibitem{selopal:etal:2020}
G.~S. Selopal, H.~Zhao, Z.~M. Wang, and F.~Rosei, ``Core/shell quantum dots solar cells,'' {\em Advanced Functional Materials}, vol.~30, no.~13, p.~1908762, 2020.

\bibitem{talapin:etal:2010}
D.~V. Talapin, J.-S. Lee, M.~V. Kovalenko, and E.~V. Shevchenko, ``{Prospects of Colloidal Nanocrystals for Electronic and Optoelectronic Applications},'' {\em Chemical Reviews}, vol.~110, no.~1, pp.~389--458, 2010.

\bibitem{shirasaki:etal:2013}
Y.~Shirasaki, G.~J. Supran, M.~G. Bawendi, and V.~Bulovi{\'c}, ``Emergence of colloidal quantum-dot light-emitting technologies,'' {\em Nature Photonics}, vol.~7, no.~1, pp.~13--23, 2013.

\bibitem{khordad:etal:2011}
R.~Khordad, G.~Rezaei, B.~Vaseghi, F.~Taghizadeh, and H.~A. Kenary, ``Study of optical properties in a cubic quantum dot,'' {\em Optical and Quantum Electronics}, vol.~42, no.~9, pp.~587--600, 2011.

\bibitem{hassanabadi:rajabi:2009}
H.~Hassanabadi and A.~A. Rajabi, ``Energy levels of a spherical quantum dot in a confining potential,'' {\em Physics Letters A}, vol.~373, no.~6, pp.~679--681, 2009.

\bibitem{kasapoglu:etal:2010}
E.~Kasapoglu, F.~Ungan, H.~Sari, and I.~S{\"o}kmen, ``The hydrostatic pressure and temperature effects on donor impurities in cylindrical quantum wire under the magnetic field,'' {\em Physica E: Low-dimensional Systems and Nanostructures}, vol.~42, no.~5, pp.~1623--1626, 2010.

\bibitem{atayan:etal:2008}
A.~K. Atayan, E.~M. Kazaryan, A.~V. Meliksetyan, and H.~A. Sarkisyan, ``Magneto-absorption in cylindrical quantum dots,'' {\em The European Physical Journal B}, vol.~63, no.~4, pp.~485--492, 2008.

\bibitem{zeng:etal:2013}
Z.~Zeng, C.~S. Garoufalis, A.~F. Terzis, and S.~Baskoutas, ``Linear and nonlinear optical properties of {ZnO/ZnS} and {ZnS/ZnO} core shell quantum dots: {E}ffects of shell thickness, impurity, and dielectric environment,'' {\em Journal of Applied Physics}, vol.~114, no.~2, 2013.

\bibitem{vasudevan:etal:2015}
D.~Vasudevan, R.~R. Gaddam, A.~Trinchi, and I.~Cole, ``Core--shell quantum dots: {P}roperties and applications,'' {\em Journal of Alloys and Compounds}, vol.~636, pp.~395--404, 2015.

\bibitem{kuo:chang:2003}
D.~M.-T. Kuo and Y.-C. Chang, ``Effects of coulomb blockade on the photocurrent in quantum dot infrared photodetectors,'' {\em Physical Review B}, vol.~67, no.~3, p.~035313, 2003.

\bibitem{wen-fang:2006}
X.~Wen-Fang, ``{Singlet-Triplet Transitions of a P\"oschl-Teller Quantum Dot},'' {\em Communications in Theoretical Physics}, vol.~46, no.~6, p.~1101, 2006.

\bibitem{mora-ramos:barseghyan:duque:2010}
M.~E. Mora-Ramos, M.~G. Barseghyan, and C.~A. Duque, ``{Excitons in cylindrical GaAs P\"oschl-Teller quantum dots: Hydrostatic pressure and temperature effects},'' {\em Physica E: Low-dimensional Systems and Nanostructures}, vol.~43, no.~1, pp.~338--344, 2010.

\bibitem{hayrapetyan:kazaryan:tevosyan:2013}
D.~B. Hayrapetyan, E.~M. Kazaryan, and H.~K. Tevosyan, ``{Optical properties of spherical quantum dot with modified P\"oschl-Teller potential},'' {\em Superlattices and Microstructures}, vol.~64, pp.~204--212, 2013.

\bibitem{hayrapetyan:kazaryan:tevosyan:2012}
D.~B. Hayrapetyan, E.~M. Kazaryan, and H.~K. Tevosyan, ``{Direct interband light absorption in the cylindrical quantum dot with modified P\"oschl-Teller potential},'' {\em Physica E: Low-dimensional Systems and Nanostructures}, vol.~46, pp.~274--278, 2012.

\bibitem{zhu:kroemer:1983}
Q.-G. Zhu and H.~Kroemer, ``Interface connection rules for effective-mass wave functions at an abrupt heterojunction between two different semiconductors,'' {\em Phys. Rev. B}, vol.~27, no.~6, p.~3519, 1983.

\bibitem{li:kuhn:1993}
T.~L. Li and K.~J. Kuhn, ``Band-offset ratio dependence on the effective-mass {H}amiltonian based on a modified profile of the $\text{GaAs-Al}_x\text{Ga}_{1-x}\text{As}$ quantum well,'' {\em Phys. Rev. B}, vol.~47, no.~19, p.~12760, 1993.

\bibitem{mustafa:mazharimousavi:2007}
O.~Mustafa and S.~H. Mazharimousavi, ``Ordering ambiguity revisited via position dependent mass pseudo-momentum operators,'' {\em Int. J. Theor. Phys.}, vol.~46, no.~7, pp.~1786--1796, 2007.

\bibitem{fuda:1969}
M.~G. Fuda, ``Hard-core potentials and the three-body problem,'' {\em Physical Review}, vol.~178, no.~4, p.~1682, 1969.

\bibitem{downs:ram:1978}
B.~W. Downs and B.~Ram, ``Hard-core potentials in quantum mechanics,'' {\em American Journal of Physics}, vol.~46, no.~2, pp.~164--168, 1978.

\bibitem{krane:1988}
K.~S. Krane, {\em Introductory Nuclear Physics}.
\newblock John Wiley, 2nd.~ed., 1988.

\bibitem{cervantes:benavides:del-rio:2007}
L.~Cervantes, A.~Benavides, and F.~Del~R{\'\i}o, ``Theoretical prediction of multiple fluid-fluid transitions in monocomponent fluids,'' {\em The Journal of chemical physics}, vol.~126, no.~8, 2007.

\bibitem{burkhardt:1968}
T.~W. Burkhardt, ``Ground-state and low-excited properties of liquid 3he calculated with a two-body potential,'' {\em Annals of Physics}, vol.~47, no.~3, pp.~516--564, 1968.

\bibitem{baker:etal:1982a}
G.~A. Baker~Jr, M.~de~Llano, J.~Pineda, and W.~C. Stwalley, ``Redetermination of hard-core square-well-potential parameters for helium using new constructive methods for the ground state of liquid \({}^4\text{He}\),'' {\em Physical Review B}, vol.~25, no.~1, p.~481, 1982.

\bibitem{baker:etal:1982b}
G.~A. Baker~Jr, L.~Benofy, M.~Fortes, M.~de~Llano, S.~Peltier, and A.~Plastino, ``Hard-core square-well fermions,'' {\em Physical Review A}, vol.~26, no.~6, p.~3575, 1982.

\bibitem{dong:lozada-cassou:2005}
S.-H. Dong and M.~Lozada-Cassou, ``Exact solutions of the {S}chr{\"o}dinger equation with the position-dependent mass for a hard-core potential,'' {\em Physics Letters A}, vol.~337, no.~4-6, pp.~313--320, 2005.

\bibitem{shankar:2013}
R.~Shankar, {\em Principles of Quantum Mechanics}.
\newblock Springer, 2nd~ed., 2013.

\bibitem{khlevniuk:tymchyshyn:2018}
A.~Khlevniuk and V.~Tymchyshyn, ``Classical treatment of particle with position-dependent mass $m (r)= 1/(1+ r^4)$ in 1{D} and 2{D} subjected to harmonic potential,'' {\em Journal of Mathematical Physics}, vol.~59, no.~8, 2018.

\bibitem{morrow:brownstein:1984}
R.~A. Morrow and K.~R. Brownstein, ``Model effective-mass {H}amiltonians for abrupt heterojunctions and the associated wave-function-matching conditions,'' {\em Phys. Rev. B}, vol.~30, no.~2, p.~678, 1984.

\bibitem{karabulut:baskoutas:2008}
{\.I}.~Karabulut and S.~Baskoutas, ``Linear and nonlinear optical absorption coefficients and refractive index changes in spherical quantum dots: Effects of impurities, electric field, size, and optical intensity,'' {\em Journal of Applied Physics}, vol.~103, no.~7, 2008.

\bibitem{xie:2010}
W.~Xie, ``Impurity effects on optical property of a spherical quantum dot in the presence of an electric field,'' {\em Physica B: Condensed Matter}, vol.~405, no.~16, pp.~3436--3440, 2010.

\bibitem{keshavarz:zamani:2013}
A.~Keshavarz and N.~Zamani, ``Optical properties of spherical quantum dot with position-dependent effective mass,'' {\em Superlattices and microstructures}, vol.~58, pp.~191--197, 2013.

\bibitem{miranda:mora-ramos:duque:2013}
G.~Miranda, M.~E. Mora-Ramos, and C.~A. Duque, ``{Exciton-related nonlinear optical absorption and refraction index change in GaAs-Ga1-xAlxAs double quantum wells},'' {\em Physica B}, vol.~409, pp.~78--82, 2013.

\bibitem{kasapoglu:etal:2022}
E.~Kasapoglu, M.~B. Y{\"u}cel, S.~Sakiroglu, H.~Sari, and C.~A. Duque, ``{Optical Properties of Cylindrical Quantum Dots with Hyperbolic-Type Axial Potential under Applied Electric Field},'' {\em Nanomaterials}, vol.~12, no.~19, p.~3367, 2022.

\bibitem{yucel:kasapoglu:duque:2022}
M.~B. Y{\"u}cel, E.~Kasapoglu, and C.~A. Duque, ``Effects of intense laser field on electronic and optical properties of harmonic and variable degree anharmonic oscillators,'' {\em Nanomaterials}, vol.~12, no.~10, p.~1620, 2022.

\bibitem{bayrak:kaya:bayrak:2023}
K.~Bayrak, D.~Kaya, and O.~Bayrak, ``The effect of position-dependent effective mass on the optical properties of a spherical quantum dot confined in inverse square root truncated and deformed exponential potential,'' {\em The European Physical Journal Plus}, vol.~138, no.~6, pp.~1--11, 2023.

\bibitem{sakiroglu:yucel:kasapoglu:2023}
{Sakiroglu, S.}, {Y\"ucel, M. B.}, and {Kasapoglu, E.}, ``The effects of the variable mass on the electronic and nonlinear optical properties of octic anharmonic oscillators,'' {\em Eur. Phys. J. Plus}, vol.~138, no.~10, p.~946, 2023.

\bibitem{naimi:vahedi:soltani:2015}
Y.~Naimi, J.~Vahedi, and M.~Soltani, ``Effect of position-dependent effective mass on optical properties of spherical nanostructures,'' {\em Optical and Quantum Electronics}, vol.~47, pp.~2947--2956, 2015.

\end{thebibliography}

\end{document}